\documentclass[twocolumn]{article}

\usepackage{graphicx}
\usepackage{amssymb,amsfonts,amsmath}

\begin{document}
\onecolumn
\title{Collective and single cell behavior in epithelial contact inhibition}
\author{Alberto Puliafito\thanks{Kavli Institute for Theoretical
    Physics,  UCSB, Santa Barbara, CA, 93106, USA}  \thanks{co-first author}, \
 Lars Hufnagel\thanks{European Molecular Biology Laboratory,
   Meyerhofstra\ss e 1, 69117 Heidelberg, Germany}  \footnotemark[2], \
  Pierre Neveu\footnotemark[1], \ Sebastian Streichan\footnotemark[3],\\
  Alex Sigal\thanks{Division of Biology, California Institute of Technology, Pasadena, CA 91125, USA}, \
  Deborah K. Fygenson\thanks{Department of Physics and Program in Biomolecular Science \& Engineering, UCSB, Santa Barbara, CA, 93106, USA} \ and
 Boris I. Shraiman\footnotemark[1]  \footnotemark[5]}

\maketitle

\begin{abstract}
Control of cell proliferation is a fundamental aspect of tissue physiology central to morphogenesis,  wound healing and cancer. 
Although many of the molecular genetic factors are now known, the system level regulation of growth 
is still poorly understood. A simple form of inhibition of cell proliferation is encountered \textit{in vitro} in  normally differentiating epithelial cell cultures  
and is known as "contact inhibition".  The study presented here provides a quantitative characterization of contact inhibition dynamics on tissue-wide and single cell levels. Using long-term tracking of cultured MDCK  cells we demonstrate that inhibition of cell division in a confluent monolayer follows inhibition of cell motility and sets in when mechanical constraint on local expansion causes divisions to reduce cell area. We quantify cell motility and cell cycle statistics in the low density confluent regime and their change across the transition to epithelial morphology which occurs with increasing cell density. We then study the dynamics of cell area distribution arising through reductive division, determine the average mitotic rate as a function of cell size and demonstrate that complete arrest of mitosis occurs when cell area falls below a critical value. We also present a simple computational model of growth mechanics which captures  all aspects of the observed behavior. 
Our measurements and analysis show that contact inhibition is a consequence of mechanical interaction and constraint rather than interfacial contact alone, and 
define quantitative phenotypes that can guide future studies of molecular mechanisms underlying contact inhibition. 
\end{abstract}
\twocolumn
\section{Introduction}
The precise orchestration of cell division and growth is central to morphogenesis and animal development~\cite{Leevers2005,Lecuit2007}. Complex  cellular signaling and regulatory networks  are dedicated to  growth control  and misregulation of cell proliferation
 leads to  tumors and  cancer~\cite{Alberts08}.
Epithelial tissue is 
an important system to study regulation of growth.  Normal development of epithelial tissue involves a mesenchymal to epithelial transition (MET)~\cite{Chaffer2007} associated with the loss of cell mobility, mitotic arrest and acquisition of epithelial morphology. This transition is reversed in the process of wound healing~\cite{Guarino1999}. On the other hand, cells that have undergone oncogenic epithelial to mesenchymal transition (EMT) typically lose their ability to undergo MET. Hence understanding the normal MET process is of fundamental importance for understanding oncogenic transformations which disregulate it.

In cultured, non-cancerous epithelial cells, the transition from freely proliferating, non-confluent cells to 
fully differentiated, dense epithelial monolayers is commonly referred to as ``contact inhibition''~\cite{Abercrombie1967,Castor1968,Abercrombie1970,Martz1972}. Contact inhibition in confluent cell cultures 
is currently defined as i) a dramatic decrease of cell mobility and mitotic rate with increasing cell density; ii) establishment of a stationary post-confluent state which is insensitive to nutrient renewal. It is widely believed that contact inhibition, as the name suggests, is caused by cell contact. But despite extensive study, current understanding of the mechanism of contact inhibition is far from complete (see~\cite{Huttenlocher1998,Halbleib2006,Takai2008,Zeng2008,Heckman2009}).\\ 

Many molecular mechanisms have been proposed to contribute to contact inhibition. It is widely accepted that contact inhibition requires establishment of E-cadherin mediated cell-cell contacts 
and subsequent maturation of the adherens junctions (AJs) that link E-cadherin and F-actin in a synapse-like complex involving numerous other proteins~\cite{Jamora2002,Yamada2005,Tamada2007,Tinkle2008}. However, the nature of the signaling pathway leading to suppression of mitosis remains unclear. One possible pathway involves $\beta$-catenin, a  mediator of Wnt signaling, that, in addition to its function as a transcriptional co-factor, is  associated with the AJs at the cell surface~\cite{Orsulic1999,Gottardi2001a}. A contact inhibition role has been reported for NF2/Merlin, a tumor suppressor gene~\cite{Hamaratoglu2006,Curto07} that encodes a  membrane-cytoskeletal scaffolding protein, which most likely acts via the Hippo kinase pathway, controlling nuclear localization of the transcriptional activator YAP~\cite{Zhao2007,Pan07,Zeng2008} - itself a known regulator of cell proliferation. Contact inhibition is known to involve the  MAPK pathway, which, in 
turn, promotes cell cycle entry by regulating the expression of cyclinD1~\cite{Shixiong2004,Matsubayashi2004,Fournier2008}. Also implicated are Nectins~\cite{Minami2007,Sakisaka2007,Takai2008} - a family of cell adhesion molecules that are involved, together with integrins and other proteins, in the regulation of cell motility and proliferation. Yet, this accumulated knowledge falls far short of a comprehensive picture of contact inhibition. The difficulty in achieving a better understanding of the molecular mechanism lies in the complexity of the contact inhibition phenotype
, which, as we describe below, involves the concurrence of many processes.

To facilitate  progress in the dissection of the regulatory pathways underlying contact inhibition, we undertook a quantitative reexamination of  the spatio-temporal dynamics of an adherent epithelial layer formed by Madin-Darby Canine Kidney (MDCK) cells.  These cells are known for their ability to 
exhibit contact inhibition and achieve characteristic epithelial morphology in culture~\cite{Gaush1966,Rothen-Rutishauser1998} thus providing an excellent model system for {\it in vitro} study of epithelial tissue dynamics ~\cite{Poujade2007,Petitjean2010,Angelini2011}. Using long-term fluorescence and phase contrast video microscopy in conjunction with image segmentation and cell tracking, we have characterized the temporal progression of contact inhibition in 
growing MDCK colonies. Quantitative analysis of the evolution of cell density, cell motility and cell division rate reveals that contact inhibition proceeds in three distinct stages: 1) 
a stage of cell density growth with gradual inhibition of motility, but without inhibition of mitosis that is followed by 2) a rapid transition to epithelial cell morphology, followed by 3) continued cell division and reduction of cell size with a progressively decreasing rate of mitosis. Mitotic arrest is achieved once cell area falls below a certain threshold.  Our findings 
show that contact between cells is not sufficient for inhibition of 
mitosis in MDCK cells. Instead, inhibition of cell proliferation is a consequence of  mechanical constraint that causes successive 
cell divisions to reduce cell area.


\section{Results}
\subsection{Large scale analysis}
To separate the effect of cell contact from that of mechanical constraint arising upon confluence of proliferating cells, we first examine the dynamics of 
isolated, growing colonies of MDCK cells. The colonies, started from a small initial number 
of cells, were monitored with subcellular resolution by time-lapse video microscopy for up to three weeks until nearly complete proliferation arrest 
 (see Materials and Methods 
 for details).
Fig. 1A shows the large scale dynamics of a growing colony. 
The 
boundary of the colony exhibits 
nontrivial dynamics due to the combined effect of motility and cell division. 
It 
moves 
outward with a non-uniform velocity forming finger-like protrusions~\cite{Poujade2007,Mark2010,Petitjean2010}.
Yet the total area of the colony grows following a simple exponential law  (Fig. 1B) for up to 5-6 days, reaching over $10^3$ cells. 
Cell density in the bulk  remains constant during this period (
Fig. 1C). Daughter cells 
occupy, on average, twice the area of their mother cell and the rate of colony area growth matches exactly the rate of cell mitosis. Thus, colony expansion is driven by cell proliferation.
In this ``free growth'' regime, although cells stay in contact with each other for several days, no inhibition of growth is observed.

Expansion of the colony is made possible by the fluid-like motion of cells in the two-dimensional confluent layer (see Fig.2A). To quantitatively characterize this motion 
we 
carried out a PIV-type analysis of the 
time series of phase-contrast  images~\cite{Raffel1998}.  
This analysis determines the local velocity field 
by comparing successive images.  In the free growth regime, cells exhibit a swirling, but outwardly biased flow with 
a root-mean-square (r.m.s.) outward velocity of about $15\mu$m/h, 
or approximately 1 cell width per hour. The motion of nearby cells is correlated on a length scale of about 5 cells (see Fig.2C). 

Of course, exponential increase in colony area  cannot continue indefinitely as it would require the outward motion of peripheral cells 
to have an exponentially increasing velocity. 
To support an exponential area increase 
at a rate $1/\tau$, the velocity of cells on the boundary 
must be $v_b\simeq \sqrt{A/4\pi\tau^2}$. Comparing this to the observed maximal velocity of cell motion, $v_{\rm max} \simeq 15\mu$m/h, we arrive at 
an estimate for the critical size of the colony, $A_c\simeq 2\cdot10^6 \mu m^2$. Above this critical area, 
expansion of the colony cannot keep up with cell proliferation in the bulk without increasing cell density. This estimate is close to the observed area at the time of crossover (
$\sim6$ days) from exponential to sub-exponential growth of the colony area (Fig.1B). The crossover is indeed 
coincident with the onset of a gradual increase in cell density in the bulk of the colony, as shown in Fig. 1C. Single cell analysis 
(below) confirms that mitotic rate in this "pre-transition" regime does not decrease so that 
 the sub-exponential expansion of the colony area is accounted for by increasing density alone.  

Thus, it appears that 
increased cell density (and the associated 
decrease of average cell size) 
is a consequence of mechanical constraint imposed by the inability of the tissue at the periphery to expand fast enough to accommodate cell proliferation in the bulk. 
 As cell density begins to increase, cell motility starts to decrease as shown in Fig. 2C (see also ref.~\cite{Petitjean2010}). The correlation length associated 
 with the velocity field exhibits a 
 peak which closely corresponds to the  transition to epithelial cell morphology, which we shall define and discuss in detail below. 
The correlation length of motion subsequently decreases with time down to the size of  a single cell (where the displacement is comparable to the optical resolution of the images), indicating that large scale swirls observed in the 
 free growth regime are 
 disappearing and cell motion is reducing to small scale fluctuations. The observed decrease in the root-mean-square velocity and the correlation length of cell motion can be understood in terms of a rapid increase in cell-substrate adhesion in the process of morphological transition. 

Cell behavior in the center of a colony at the end of the free 
growth regime is similar to what is observed in 
confluent cell cultures that were seeded homogeneously, 
see Fig.2C,D.
In homogeneously seeded cultures the space constraint is more severe and cell density increases more rapidly upon confluence, reaching the morphological 
transition soon after initiation of the culture. We note that panels C and D in Fig. 2 
differ also in the substrate: glass and PDMS  respectively (see Materials and Methods). 
Although the magnitude of the correlation length 
differs 
in the two cases, the qualitative behavior and the 
morphological transition are very similar.

To investigate the long term evolution of cell size and density upon exit from the free growth regime
we measured  the average cell area (using digital image segmentation) over a period of 20 days.  As shown in Fig.3C, 
average cell area decreases 
10-fold over a 15 day period. In the same period, average cell height increases 
 only by a factor of 2 (from 5-6 $\mu$m to 12-15 $\mu$m
, see Supplementary Information Fig.S1D) thus indicating that cell volume decreases.
Fig. 3C 
shows a rather sharp crossover from rapid to slow growth of cell density (and corresponding decrease of cell size). This crossover coincides with the  transition in cell morphology illustrated by the comparison of panels A and B in Fig.3 
and separates the "pre-transition" and "post-transition" stages of the contact inhibition process. The pre-transition transient is characterized by the gradual reduction of cell motility (Fig. 2
) discussed above. The post-transition state  is characterized by the
absence of cell rearrangement, except through cell division. Mitotic rate decreases continuously in the post-transition stage (see Single cell analysis section) leading to arrest of cell proliferation independent of nutrient renewal. This ``arrested'' 
 regime can last for weeks.  (We monitored the tissue for 23 days without detecting any significant changes in the area distribution.) 
However, the state of proliferation arrest can be readily reversed by scratching the cell layer to create a free boundary~\cite{Poujade2007}, or by stretching the substrate along with the cell layer. 


The morphological transition itself is readily quantified by the radial distribution function which measures conditional probability of finding a cell at a given distance from a reference cell (Fig. 3D
). In the pre-transition stage, the distribution function exhibits an exclusion zone at distances comparable to the size of the nucleus, and is flat for larger distances, indicating a disordered system of non-overlapping cells. In the post-transition regime 
a peak and a trough emerge in the distribution function, corresponding to 
nearest and next-nearest neighbors. This indicates an increase in size homogeneity and the appearance of local ordering of cells within the tissue. 
Further quantitative characterization of the these regimes is provided by single cell tracking and analysis. 

\subsection{Single cell analysis}
To further quantify cell behavior in the different regimes of tissue growth, we followed individual cells 
through the division process. 
In the free growth regime, 
each daughter cell grows back to the size of 
its mother cell, as shown in Fig.~\ref{fig:single}A.
In contrast, in both pre- and post- transition regimes mitosis reduces  cell area  
by approximately a factor of two without subsequent growth of daughter cells 
(
Fig~\ref{fig:single}A, B).  The prevalence of reductive division is demonstrated in Fig.~\ref{fig:single}C, which compares measured areas of mother and daughter cells. 
Combined area of daughter cells does not exceed the area of the "mother" cell, independent of the area of the latter
(Fig.~\ref{fig:single}C, inset). Thus, cells in both the 
pre- and post-transition regimes are 
"incompressible" in the sense that new cells introduced through cell division do not achieve any expansion of their area at the expense of their neighbors.
 
The morphological transition coincides with 
an approximately 5-fold decrease in the average mitotic rate and a dramatic broadening 
of the distribution of cell cycle periods 
(Fig.~\ref{fig:single}E). 
Once the cell division time 
becomes significantly longer than a day, measuring mitotic rate by tracking individual cells becomes very difficult, necessitating 
a different strategy for measuring the dependence of 
mitotic rate on cell size. The latter, as we show next, can be deduced from a quantitative study of the temporal evolution of the cell area distribution.
 
To measure the distribution of cell area as a function of time in the post-transition regime, we 
performed a computerized segmentation of fluorescent images (using the MDCK-EcadGFP cell line) (
Fig. 3B 
and 
Fig.~\ref{fig:single}D). Over 
a period of about 15 days following the morphological transition the average cell area decreases 6-fold.  Cell size 
converges on a narrow, stationary size distribution centered 
about an average area of 35$\mu$m$^2$ (see SI, Fig.~S1 for more details on cell morphology).  
%

Because cell area in the post-transition phase 
is approximately constant between successive divisions (
Fig.~\ref{fig:single}A-B), the dynamics of the cell area distribution is due solely to mitosis. (Rates of apoptosis, about 0.02 per day per cell, are negligible by comparison with mitosis
.)  Thus the difference in area distributions at two consecutive time points reflects loss of larger cells which upon division give rise each to a pair of cells at approximately half the size as represented by: 
\begin{equation}\label{eq:master}
\partial_t n(a,t) = 2 \gamma (2a) n(2a,t)-\gamma (a) n(a,t)
\end{equation}
where $n(a,t)$ denotes the expected number of cells with area, $a$, at time $t$ and $\gamma (a)$ represents the average rate of division as a function of cell area. 
Fitting the observed temporal changes in cell area distribution to Eq.~\eqref{eq:master} allows us to determine the mitotic rate $\gamma (a)$.  The dependence of mitotic rate on cell size is shown in Fig.~\ref{fig:single}F.
The result is consistent with the conclusion made on the basis of  the single cell measurements, Fig.~\ref{fig:single}E: a rapid decrease in the mitotic rate
once cell area falls below critical, which fits approximately the Hill function form:   $\gamma(a)/ \gamma_0=a^m / (a^m + a_0^m)$  with $m \approx 4$ and $a_0\approx 170\mu m^2$, where $\gamma_0$ is the  division rate in the free growth regime. The observed reductive nature of cell division  and the dependence of mitotic rate on cell size together explain the dynamics of tissue density in the post-transition regime and the convergence towards proliferation arrest, thus capturing the dynamics of the "contact inhibition" process.

\subsection{A model of self-limiting growth of adherent cell monolayer} 
To illustrate our interpretation of the observed interplay between cellular growth, motility and colony expansion, we formulate and analyze a simple growth model for adherent epithelial tissues. We choose as a point of departure a one dimensional version of the "vertex model" \cite{Hufnagel07,Julicher2007} as depicted in Fig. 5A.
The details of the model are described in the supplemental material. Briefly, we assume that cells, specified by their vertices $r_i$ and $r_{i+1}$, form a connected tissue. The short time elastic response of cells has a Hooke's law dependence (see SI Eq 2.1) on the difference between the current length of the cell, $l_i=|r_{i+1}-r_i|$ and the  intrinsic preferred length $L_i$. In addition cells interact with the substrate. To represent the effect of substrate adhesion and of cell motility we introduce for each cell an "attachment point" $R_i$, connected to the cell by a spring, and endowed with relaxational dynamics with friction $\sigma$ and random Langevin driving force $\eta_i(t)$ (see SI Eq 2.2). This (gaussian white) random force  represents cell crawling, its variance $\Gamma$ defining motility. The force is assumed to average out to zero in the bulk, but not on the boundary where to represent the outward bias of the boundary cell motion  we allow $<\eta_b> = \sigma v_{max}$, where  $v_{max}$ sets the maximal velocity.

The model also includes cell growth and proliferation. Cell growth is represented by allowing intrinsic cell size $L_i$ to increase with time. However, motivated by our experimental finding that cells in dense tissues don't actively push on their neighbors, we allow $L_i$ to grow only if the cell is stretched by the surrounding tissue. Because stretching corresponds to $\delta l_i=l_i -L_i >0$ we take $dL_i /dt$ to be a simple step function of $\delta l_i$ with the threshold at zero. Cell division splits a cell into two, with intrinsic size of each daughter equal to $L_i/2$ of the mother. Guided by our experimental observations  (Fig.~\ref{fig:single}) we make the rate of cell division explicitly dependent on cell size $l_i$ via $p(l_i)={\rm max}(\gamma(l_i-l_{\rm min}),0)$ which implements a "size check point" ($l_i>l_{\rm min}$) for cell proliferation. 

The spatio-temporal growth dynamics of an initially small colony is shown in Fig. 5B. 
Independent of the specific parameters used in our simulations, we find two growth regimes: an initial exponential growth of the colony with uniform proliferation is followed by a linear growth regime with cell proliferation mainly on the margin of the colony. The maximal size of a freely proliferating colony and the critical time for cross over from exponential to sub-exponential growth may be obtained by the following argument. A  patch consisting of $N(t)=e^{\gamma t}$ proliferating cells increases its diameter with the instantaneous speed ${l_0 d  {N(t)} /dt}$. This speed cannot exceed the maximal velocity of the interface $v_{max}$,  lest proliferating cells become compressed, as the colony becomes unable to spread fast enough to keep up with cell proliferation. The two velocities became equal at $t_c =  \gamma^{-1} \log (2 v_{max} /\gamma l_0)$ at which time the colony size is  $2v_{max}/\gamma$. (This 1D argument is readily generalized to 2D.)

The two growth regimes are manifest on the colony scale but are also reflected on a single cell level (Fig. 5B inset). Large cells on colony margins re-grow in size after division and then divide again. In higher density regions, cells undergo size reductive divisions until the division finally ceases. The temporal evolution of the cell sizes in the middle of the colony is shown in Fig. 5C. 
The difference in cell behavior arises, in our model, from the difference in their mechanical state. Motile cells in small colonies or on the periphery of a large colony are under tension and grow in size after division. In contrast, cells in the bulk of a large colony are "boxed in" by their neighbors, and after each division reduce in size until they divide no more. Our assumptions that a) cells grow in size only under tension and b) cells do not exert compressive forces on each other, are closely related and result in a distribution of tensile stress across the tissue layer (see Figure 5D) that is consistent with the observations of Trepat et al \cite{Trepat2009}.

\section{Discussion}

Quantitative observations of cell size, motion and division rate reported here 
help dissect the complex nature of  the "contact inhibition" phenotype. 
They reveal, for example, that free exponential growth can take place within cell colonies 
even after cells have been in contact for several days. 
Such a long delay between confluence and mitotic inhibition is particular to isolated, expanding colonies.  Onset of mitotic inhibition occurs sooner in cell cultures seeded at uniform density, where 
confluence coincides with near complete occupation of the available area. 
We conclude that cell-cell contact is a necessary~\cite{Nelson2002} but not sufficient condition for growth inhibition.  

The data strongly suggests that  inhibition of cell division 
follows the reduction in cell area 
imposed by mechanical constraints 
on tissue expansion
~\cite{Chen1997,Nelson2002}. 
Interestingly, although average cell area starts to decrease in the 
motile pre-transition regime, the dependence of mitotic rate on cell area does not appear until the 
cells enter the static 
post-transition phase, 
at which point the rate of cell proliferation drops sharply. 
The cessation of cell motility and drop in mitotic rate coincide with a dramatic change in tissue morphology, as revealed by the radial distribution function of cells. The radial distribution function is thus an interesting, quantitative characterization of the state of the tissue  and its abrupt change can serve to pin-point the morphological transition, which may be considered a key element of the mesenchymal-to-epithelial transition (MET).  

Our study focused on the dependence of mitotic rate on cell area because the latter is directly measurable. Yet it is important to emphasize that the real trigger of intracellular signals 
responsible for the suppression of motility,  inhibition of mitosis, and MET may be not cell size {\em per se}, but mechanical stress and deformation which are known to induce reorganization of focal adhesion, adherens junctions, and cytoskeleton~\cite{Huttenlocher1998,Yeung2005,Pugacheva2006,Margadant2007,Janmey2007}. 

It is widely believed that mechanical tension promotes cell division~\cite{Huang1999,Schwartz2001}.  Recent direct measurements by Trepat and coworkers~\cite{Trepat2009} demonstrate that motile MDCK cells exert inward pointing traction on the substrate, implying that 
spreading cell colonies are under tension.  These measurements correspond to MDCK colonies in what we refer to as the free growth regime. Combining 
their observations with ours suggests that 
the 
motile pre-transition regime and the morphological transition itself 
correspond to the gradual relief of tension and the onset of compression brought on by cell proliferation. It has 
been suggested (in the context of the problem of organ size determination) that mechanical compression may be providing an inhibitory signal for mitosis~\cite{Shraiman2005,Hufnagel07,Aegerter-Wilmsen2007,Rauzi2008}. 
The same hypothesis could then explain the observed inhibition of mitosis in the post-transition regime. 
If so, the trigger of 
MET would be the change from tensile to compressive stress acting locally within the cell layer. Our 1D model illustrates how this scenario can generate the observed behavior. %


Our measurements also suggest that inhibition of cell division is a distinct single cell state rather than a global state induced by cell-cell signaling across the layer, as illustrated in Fig.~\ref{fig:single}E. In fact, confluent MDCK cell cultures 
with an {\it average} cell density corresponding to the morphological transition are often sufficiently heterogeneous in local cell density that highly motile cells and completely arrested cells coexist in the same colony.  Thus contact inhibition is a 
local phenomenon, which calls into question the reliability of "bulk assays" of the phenotype.
 Hence future experimental efforts focusing on molecular 
 mechanisms underlying 
 mitotic inhibition should be conducted with methods allowing single cell resolution. Such experiments, combined with 
 techniques allowing {\it in situ} measurements of mechanical stress acting on cells would, we believe, 
finally lead to decisive understanding of the contact inhibition phenomenon. 

\section{Materials and Methods}
{ \bf Cell culture}
MDCK-II cells and MDCK-Ecad-GFP were a gift from J. Nelson. Cells were cultured in
MEM (GIBCO, 11095-098) supplemented with Penicillin-Streptomycin and 5\% FBS (Cellgro, 35-010-CV) at 37$^\circ$C and 5\% CO$_2$. \\
{\bf Time-lapse microscopy}
All imaging was performed on an inverted microscope (Olympus IX-70)  with a 20X/0.7NA/Ph2 objective.
Phase and fluorescence images were taken with, respectively, a halogen lamp and an LED (Luxeon LXHL-LB5C) and
recorded to disk using a CCD camera (QImaging, Retiga EXi).  Mechanical shutters (Uniblitz VS25) in both illumination paths limited sample exposure to minimize  phototoxicity.  Stepper motors controlled the stage position and objective focus.  The shutters, stepper motors and camera were controlled by a custom-written Labview program.  Images were taken every 10 min (phase) and every 3h (fluorescence) for a given field of view.  \\
{\bf Image analysis}
Images were analyzed by using custom written Matlab programs. 
Positions of nuclei in the low cell density phase were determined from the phase contrast images,  segmented semi-automatically. 
In the static, high density regime, cells were identified by the 
fully automatic segmentation 
of Ecad-GFP fluorescence images. Mean displacement measurements were 
made by cell tracking 
using a PIV-type analysis~\cite{Raffel1998}. 
Colony profiles were obtained by extracting the boundary of the colony as a function of time 
from the images 
using standard edge detection algorithms. \\
{\bf Numerical simulations} The mathematical model underlying Fig. 5 and the method of simulation are described in the SI.

\section{Acknowledgments}
Authors acknowledge valuable interactions with T. Weimbs, M. Elowitz, D. Sprinzak, L. Peliti and M. Vergassola. AP, LH and PN were supported by NSF PHY05-51164. BIS acknowledges support of NSF PHY-08-44989.

\onecolumn


\begin{thebibliography}{10}

\bibitem{Leevers2005}
Leevers SJ, McNeill H
\newblock \emph{Controlling the size of organs and organisms.}
\newblock Curr Opin Cell Biol,  17 (2005),  pp. 604--609.

\bibitem{Lecuit2007}
Lecuit T, Le~Goff L
\newblock \emph{Orchestrating size and shape during morphogenesis.}
\newblock Nature,  450 (2007),  pp. 189--192.

\bibitem{Alberts08}
Alberts B, {et~al.}
\newblock (2008),  \emph{Molecular biology of the cell}
\newblock (Garland Science).

\bibitem{Chaffer2007}
Chaffer CL, Thompson EW, Williams ED
\newblock \emph{{Mesenchymal to epithelial transition in development and
  disease.}}
\newblock Cells Tissues Organs,  185 (2007),  pp. 7 -- 19.

\bibitem{Guarino1999}
Guarino M, Micheli P, Pallotti F, Giordano F
\newblock \emph{Pathological relevance of epithelial and mesenchymal phenotype
  plasticity}
\newblock Pathol Res Pract,  195 (1999),  pp. 379--89.

\bibitem{Abercrombie1967}
Abercrombie M
\newblock \emph{Contact inhibition: the phenomenon and its biological
  implications.}
\newblock Natl Cancer Inst Monogr,  26 (1967),  pp. 249--277.

\bibitem{Castor1968}
Castor LN
\newblock \emph{Contact regulation of cell division in an epithelial-like cell
  line}
\newblock J Cell Physiol,  72 (1968),  pp. 161--72.

\bibitem{Abercrombie1970}
Abercrombie M
\newblock \emph{Contact inhibition in tissue culture.}
\newblock In Vitro,  6 (1970),  pp. 128--142.

\bibitem{Martz1972}
Martz E, Steinberg M
\newblock \emph{The role of cell-cell contact in "contact" inhibition of cell
  division: a review and new evidence.}
\newblock J. Cell Physiol.,  79 (1972),  pp. 189--210.

\bibitem{Huttenlocher1998}
Huttenlocher A, {et~al.}
\newblock \emph{Integrin and cadherin synergy regulates contact inhibition of
  migration and motile activity}
\newblock J Cell Biol,  141 (1998),  pp. 515--26.

\bibitem{Halbleib2006}
Halbleib JM, Nelson WJ
\newblock \emph{Cadherins in development: cell adhesion, sorting, and tissue
  morphogenesis.}
\newblock Genes Dev,  20 (2006),  pp. 3199--3214.

\bibitem{Takai2008}
Takai Y, Miyoshi J, Ikeda W, Ogita H
\newblock \emph{Nectins and nectin-like molecules: roles in contact inhibition
  of cell movement and proliferation}
\newblock Nat Rev Mol Cell Biol,  9 (2008),  pp. 603--15.

\bibitem{Zeng2008}
Zeng Q, Hong W
\newblock \emph{The emerging role of the hippo pathway in cell contact
  inhibition, organ size control, and cancer development in mammals.}
\newblock Cancer Cell,  13 (2008),  pp. 188--192.

\bibitem{Heckman2009}
Heckman CA
\newblock \emph{Contact inhibition revisited}
\newblock J Cell Physiol,  220 (2009),  pp. 574--5.

\bibitem{Jamora2002}
Jamora C, Fuchs E
\newblock \emph{Intercellular adhesion, signalling and the cytoskeleton.}
\newblock Nat Cell Biol,  4 (2002),  pp. E101--E108.

\bibitem{Yamada2005}
Yamada S, Pokutta S, Drees F, Weis WI, Nelson WJ
\newblock \emph{Deconstructing the cadherin-catenin-actin complex}
\newblock Cell,  123 (2005),  pp. 889--901.

\bibitem{Tamada2007}
Tamada M, Perez TD, Nelson WJ, Sheetz MP
\newblock \emph{Two distinct modes of myosin assembly and dynamics during
  epithelial wound closure}
\newblock J Cell Biol,  176 (2007),  pp. 27--33.

\bibitem{Tinkle2008}
Tinkle CL, Pasolli HA, Stokes N, Fuchs E
\newblock \emph{New insights into cadherin function in epidermal sheet
  formation and maintenance of tissue integrity}
\newblock Proc Natl Acad Sci U S A,  105 (2008),  pp. 15405--10.

\bibitem{Orsulic1999}
Orsulic S, Huber O, Aberle H, Arnold S, Kemler R
\newblock \emph{E-cadherin binding prevents beta-catenin nuclear localization
  and beta-catenin/LEF-1-mediated transactivation.}
\newblock J Cell Sci,  112 ( Pt 8) (1999),  pp. 1237--1245.

\bibitem{Gottardi2001a}
Gottardi CJ, Wong E, Gumbiner BM
\newblock \emph{E-cadherin suppresses cellular transformation by inhibiting
  beta-catenin signaling in an adhesion-independent manner.}
\newblock J Cell Biol,  153 (2001),  pp. 1049--1060.

\bibitem{Hamaratoglu2006}
Hamaratoglu F, {et~al.}
\newblock \emph{The tumour-suppressor genes NF2/Merlin and Expanded act through
  Hippo signalling to regulate cell proliferation and apoptosis}
\newblock Nat Cell Biol,  8 (2006),  pp. 27--36.

\bibitem{Curto07}
Curto M, Cole BK, Lallemand D, Liu CH, McClatchey AI
\newblock \emph{{Contact-dependent inhibition of EGFR signaling by
  Nf2/Merlin.}}
\newblock J. Cell Biol.,  177 (2007),  pp. 893 -- 903.

\bibitem{Zhao2007}
Zhao B, {et~al.}
\newblock \emph{Inactivation of YAP oncoprotein by the Hippo pathway is
  involved in cell contact inhibition and tissue growth control.}
\newblock Genes Dev,  21 (2007),  pp. 2747--2761.



\bibitem{Pan07}
Pan D
\newblock \emph{{Hippo signaling in organ size control.}}
\newblock Genes Dev,  21 (2007),  pp. 886 -- 897.

\bibitem{Shixiong2004}
Shixiong L, Edward R, Balkovetz D
\newblock \emph{Evidence for ERK1/2 phosphorylation controlling contact
  inhibition of proliferation in Madin-Darby canine kidney epithelial cells.}
\newblock Am J Physiol Cell Physiol,  287 (2004),  pp. C432--C439.

\bibitem{Matsubayashi2004}
Matsubayashi Y, Ebisuya M, Honjoh S, Nishida E
\newblock \emph{ERK activation propagates in epithelial cell sheets and
  regulates their migration during wound healing}
\newblock Curr Biol,  14 (2004),  pp. 731--5.

\bibitem{Fournier2008}
Fournier AK, {et~al.}
\newblock \emph{Rac-dependent cyclin D1 gene expression regulated by cadherin-
  and integrin-mediated adhesion}
\newblock J Cell Sci,  121 (2008),  pp. 226--33.

\bibitem{Minami2007}
Minami Y, {et~al.}
\newblock \emph{Involvement of up-regulated Necl-5/Tage4/PVR/CD155 in the loss
  of contact inhibition in transformed NIH3T3 cells}
\newblock Biochem Biophys Res Commun,  352 (2007),  pp. 856--60.

\bibitem{Sakisaka2007}
Sakisaka T, Ikeda W, Ogita H, Fujita N, Takai Y
\newblock \emph{{The roles of nectins in cell adhesions: cooperation with other
  cell adhesion molecules and growth factor receptors.}}
\newblock Curr Opin Cell Biol,  19 (2007),  pp. 593 -- 602.

\bibitem{Gaush1966}
Gaush~C.R., Hard~W.L. ST
\newblock \emph{Characterization of an established line of canine kidney cells
  (MDCK)}
\newblock Proceedings of the Society for Experimental Biology and Medicine,
  122 (1966),  pp. 931--5.

\bibitem{Rothen-Rutishauser1998}
Rothen-Rutishauser B, Kramer SD, Braun A, Gunthert M, Wunderli-Allenspach H
\newblock \emph{MDCK cell cultures as an epithelial in vitro model:
  cytoskeleton and tight junctions as indicators for the definition of
  age-related stages by confocal microscopy.}
\newblock Pharm Res,  15 (1998),  pp. 964--971.

\bibitem{Poujade2007}
Poujade, M., {et~al.}
\newblock \emph{Collective migration of an epithelial monolayer in response to a model wound.}
\newblock Proc Natl Acad Sci U S A, 104 (2007), pp. 15988--15993.

\bibitem{Petitjean2010}
Petitjean L, {et~al.} 
\newblock \emph{Velocity fields in a collectively migrating epithelium}
\newblock Biophys J,  98 (2010), pp. 1790--1800.

\bibitem{Angelini2011}
Angelini TE, Hannezo E, Trepat X, Marquez M, Fredberg JJ, Weitz DA.
\newblock \emph{{Glass-like dynamics of collective cell migration.}}
\newblock Proc Natl Acad Sci U S A. 108 (2011), pp. 4714--9.

\bibitem{Mark2010}
Mark S, {et~al.}
\newblock \emph{Physical model of the dynamic instability in an expanding cell
  culture}
\newblock Biophys J,  98 (2010),  pp. 361--70.



\bibitem{Raffel1998}
Raffel~M., Willert Christian~E. KJ
\newblock (1998),  \emph{Particle Image Velocimetry: A Practical Guide.}
\newblock (Springer).

\bibitem{Nelson2002}
Nelson CM, Chen CS
\newblock \emph{Cell-cell signaling by direct contact increases cell
  proliferation via a PI3K-dependent signal}
\newblock FEBS Lett,  514 (2002),  pp. 238--42.

\bibitem{Chen1997}
Chen CS, Mrksich M, Huang S, Whitesides, G. M. \&~Ingber DE
\newblock \emph{Geometric control of cell life and death}
\newblock Science,  276 (1997),  pp. 1425--1428.

\bibitem{Yeung2005}
Yeung T, {et~al.}
\newblock \emph{Effects of substrate stiffness on cell morphology, cytoskeletal
  structure, and adhesion}
\newblock Cell Motil Cytoskeleton,  60 (2005),  pp. 24--34.

\bibitem{Pugacheva2006}
Pugacheva EN, Roegiers F, Golemis EA
\newblock \emph{Interdependence of cell attachment and cell cycle signaling}
\newblock Curr Opin Cell Biol,  18 (2006),  pp. 507--15.

\bibitem{Margadant2007}
Margadant C, van Opstal A, Boonstra J
\newblock \emph{Focal adhesion signaling and actin stress fibers are
  dispensable for progression through the ongoing cell cycle}
\newblock J Cell Sci,  120 (2007),  pp. 66--76.

\bibitem{Janmey2007}
Janmey PA, McCulloch CA
\newblock \emph{Cell mechanics: integrating cell responses to mechanical
  stimuli.}
\newblock Annu Rev Biomed Eng,  9 (2007),  pp. 1--34.

\bibitem{Huang1999}
Huang S, Ingber D
\newblock \emph{The structural and mechanical complexity of cell-growth
  control.}
\newblock Nat Cell Biol.,  1(5) (1999),  pp. E131--138.

\bibitem{Schwartz2001}
Schwartz MA, Assoian RK
\newblock \emph{Integrins and cell proliferation: regulation of
  cyclin-dependent kinases via cytoplasmic signaling pathways}
\newblock J Cell Sci,  114 (2001),  pp. 2553--60.

\bibitem{Trepat2009}
Trepat X, {et~al.}
\newblock \emph{Physical forces during collective cell migration}
\newblock Nature Physics,  5 (2009),  p. 426.

\bibitem{Shraiman2005}
Shraiman B
\newblock \emph{Mechanical feedback as a possible regulator of tissue growth}
\newblock Proc. of the Natl. Acad. of Sci.,  102 (2005),  pp. 3318--3323.

\bibitem{Hufnagel07}
Hufnagel L, Teleman AA, Rouault H, Cohen SM, Shraiman BI
\newblock \emph{{On the mechanism of wing size determination in fly
  development.}}
\newblock Proc. Natl. Acad. Sci. USA,  104 (2007),  pp. 3835 -- 3840.

\bibitem{Aegerter-Wilmsen2007}
Aegerter-Wilmsen T, Aegerter CM, Hafen E, Basler K
\newblock \emph{Model for the regulation of size in the wing imaginal disc of
  Drosophila}
\newblock Mech Dev,  124 (2007),  pp. 318--26.

\bibitem{Rauzi2008}
Rauzi M, Verant P, Lecuit T, Lenne PF
\newblock \emph{Nature and anisotropy of cortical forces orienting Drosophila
  tissue morphogenesis.}
\newblock Nat Cell Biol,  10 (2008),  pp. 1401--1410.

\bibitem{Julicher2007}
Farhadifar R, Roper JC, Aigouy B, Eaton S, Julicher F.
\newblock \emph{The influence of cell mechanics, cell-cell interactions, and proliferation on epithelial packing.}
\newblock Curr Biol. 17 (2007), pp. 2095-104.


\bibitem{Lee1990}
Lee YK, Rhodes WT
\newblock \emph{Nonlinear image processing by a rotating kernel transformation}
\newblock Optics Letters,  15 (1990),  pp. 1383--1385.

\bibitem{Jain1991}
Jain AK, Farrokhnia F
\newblock \emph{Unsupervised texture segmentation using Gabor filters}
\newblock Pattern Recognition,  24 (1991),  pp. 1167--1186.

\bibitem{Ben-Israel2003}
Ben-Israel A, Greville T
\newblock (2003),  \emph{Generalized inverses: theory and applications}
\newblock (Springer-Verlag).


\end{thebibliography}

\graphicspath{{./}}
\begin{figure*}
\centerline{\includegraphics[width=8.7cm]{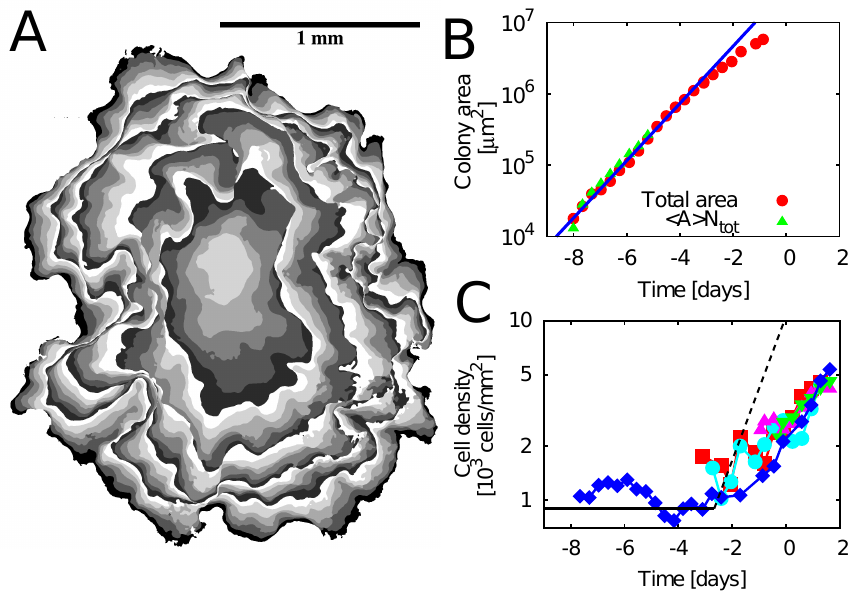}}
{\caption{{\bf Epithelial colony growth.} 
{\bf A)} Superimposed snapshots of a single colony at different times, coded by different shades of gray. Time-points were chosen to keep area increment constant. Black contours correspond to  3.0, 4.8, 5.5, 5.9, 6.3 days after seeding. 
{\bf B)} Total area of the spreading colony. Time is counted relative to the "morphological transition" at $t=0$ (see text). Green points represent total cell number (independently measured) multiplied by the average cell area.  The blue line is exponential growth with the average cell cycle time $\tau_2=0.75\pm0.14$ (s.e.m.) days (measured independently by single cell tracking).
{\bf C)} Cell density in the inner region of the colony (different colors distinguish different fields of view).  The solid black line at constant density and is a guide for the eye. The dashed black line represents exponential growth of density expected for continued cell proliferation without cell motion. 
}}
\label{fig:colony}
\end{figure*}

\begin{figure*}
\centerline{\includegraphics[width=8.7cm]{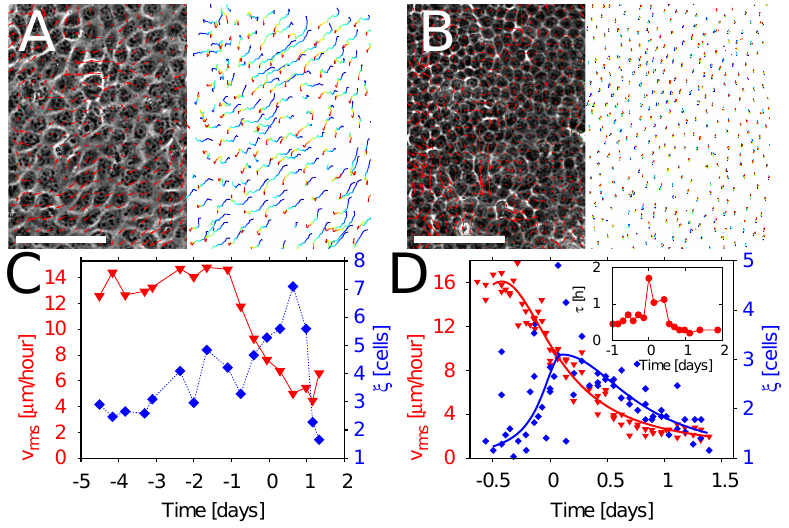}}

{\caption{{\bf Correlation analysis of cell motility.} 
{\bf A-B)}   Phase images of a confluent layer (1h before and 27 h after the morphological transition) with overlaid instantaneous velocity field (measured by PIV and interpolated) side by side with cell trajectories integrated over 200 minutes with blue and red labeling respectively the beginning and the end. 
Scale bar is 100$\mu$m. 
{\bf C-D)} R.m.s. velocity of cell motion (red symbols) and the correlation length (blue symbols) across the morphological transition in a 
in the bulk of the expanding colony (C) and
 in the continuous confluent layer plated at higher initial density (D).  Data pooled from four different 450x336 $\mu m^2$ fields of view. Lines are to guide the eye.
 {\bf D) inset:} Correlation time of cell trajectories. }
}
\label{fig:piv}
\end{figure*}

\begin{figure*}
\centerline{\includegraphics[width=8.7cm]{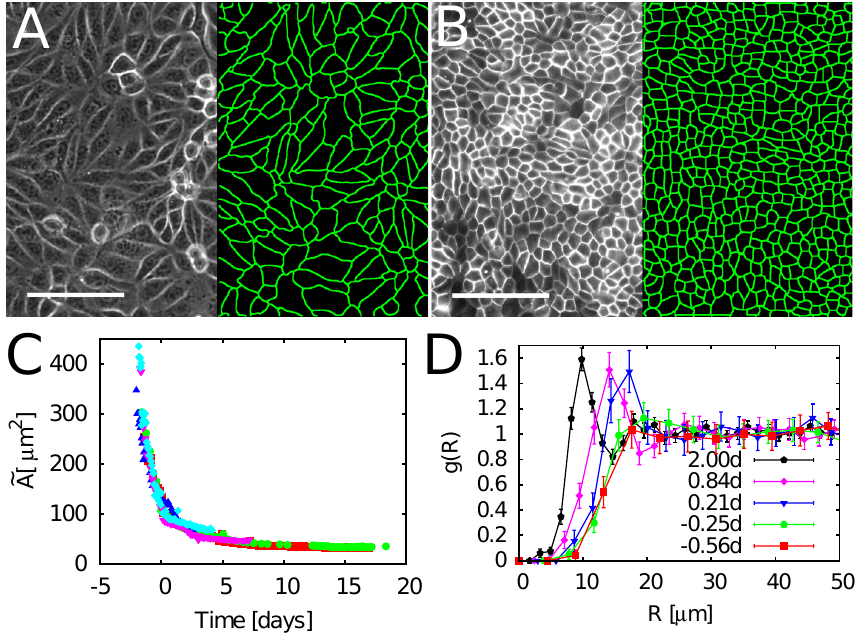}}
{\caption{{\bf Large scale quantitative characterization of contact inhibition.} {\bf A- B)} Image segmentation for MDCK cell cultures grown on PDMS. {\bf A)} phase contrast image of Ecad-GFP MDCK at low cell density. {\bf B)} fluorescent image of Ecad-GFP MDCK at high cell density. Scale bar: 100$\mu$m.  {\bf C)} Median of cell area distribution (over a 450x336 $\mu m^2$ field of view) as a function of time (t=0 set to the morphological transition). Here MDCK cells were seeded at uniform density (see methods) and imaging commenced upon confluency. 
Different line colors represent different experiments which are 
time aligned. No density change is detectible after 15-18 days. 
{\bf D)} Radial distribution function of cells at different times across the morphological transition. $g(R)$ is the ratio between the density of cells in a circular annulus distance $R$ from a reference cell and the average density. The appearance of a peak (and a trough) in the static post-transitional phase represents increased short range ordering of cells.  t=0 is defined by the first appearance of the peak: $max(g(R))>1.2$.
}}
\label{fig:avarea}
\end{figure*}

\begin{figure*}
\centerline{\includegraphics[width=8.7cm]{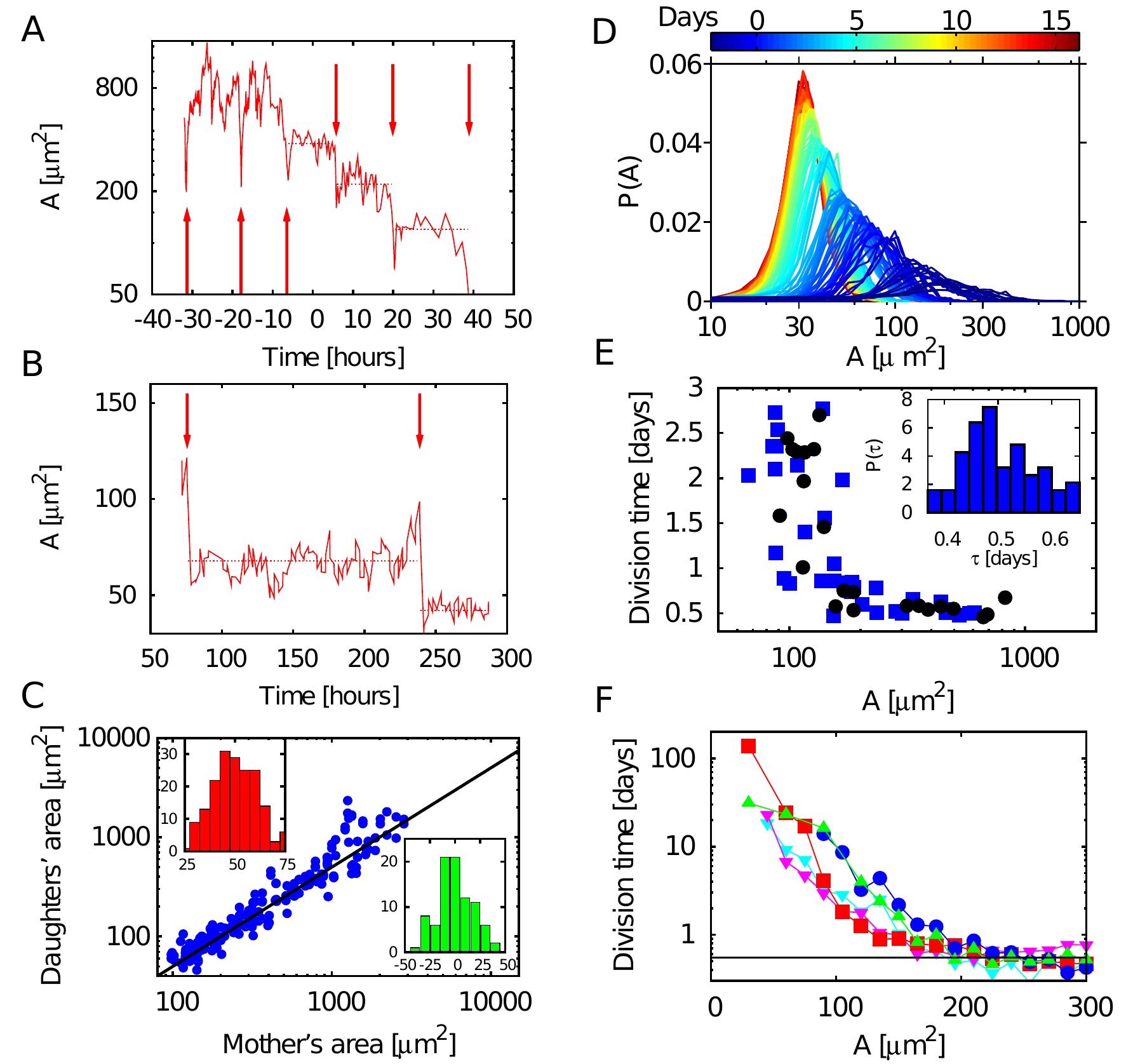}}
{\caption{{\bf Single cell level quantification of contact inhibition. } 
{\bf A)} and {\bf B)} Traces of single cell area tracked as a function of time. Arrows represent mitosis. {\bf A)}  starts below confluence and reaches high density confluence; {\bf B)} is in the post-transition phase. Dashed lines represent temporal averages.
{\bf C)} Daughter cell area versus the area of the mother cell. Data represent 96 divisions at different times for confluent layers. . Daughter cell area as the average over three time points 1 hour apart, 12 hours after mitosis. Mother vs daughter cell areas follow the line $y=x/2$, plotted in black. {\bf C) Upper inset:} Distribution of daughter cell areas in \% of the mother cell area. {\bf C) Lower inset:} Deviation of the total daughter cell area from the mother cell area.
{\bf D)} Distribution of cell area in the post-transition regime. Color codes for time. Each distribution represents the population of (at least 200) cells in the same 336x450 $\mu m^2$ field of view.
Cell area is measured by means of computer segmentation of MDCK-Ecad-GFP fluorescent images (see fig.~\ref{fig:avarea}A).
{\bf E)} Single cell division times as a function of premitotic area. Different colors represent different experiments. We note that cell cycle time increases dramatically for  cell areas below 200 $\mu m^2$. The absence of data below $70 \mu m^2$ is due to the difficulty of tracking single 
cells in that regime. {\bf E) inset:}  Distribution of division times in the pre-transition regime.
{\bf F)}  Division times as a function of cell area inferred from the dynamics of the $P(A)$ functions using Eq.~\eqref{eq:master} (see methods). Different colors represent different experiments. The black line represents average division time in the pre-transition regime. Cell division slows down for cell size  below 200$\mu m^2$, consistent with cell tracking measurements shown in panel E.}
\label{fig:single}}
\end{figure*}

\begin{figure*}
\centerline{\includegraphics[width=8.7cm]{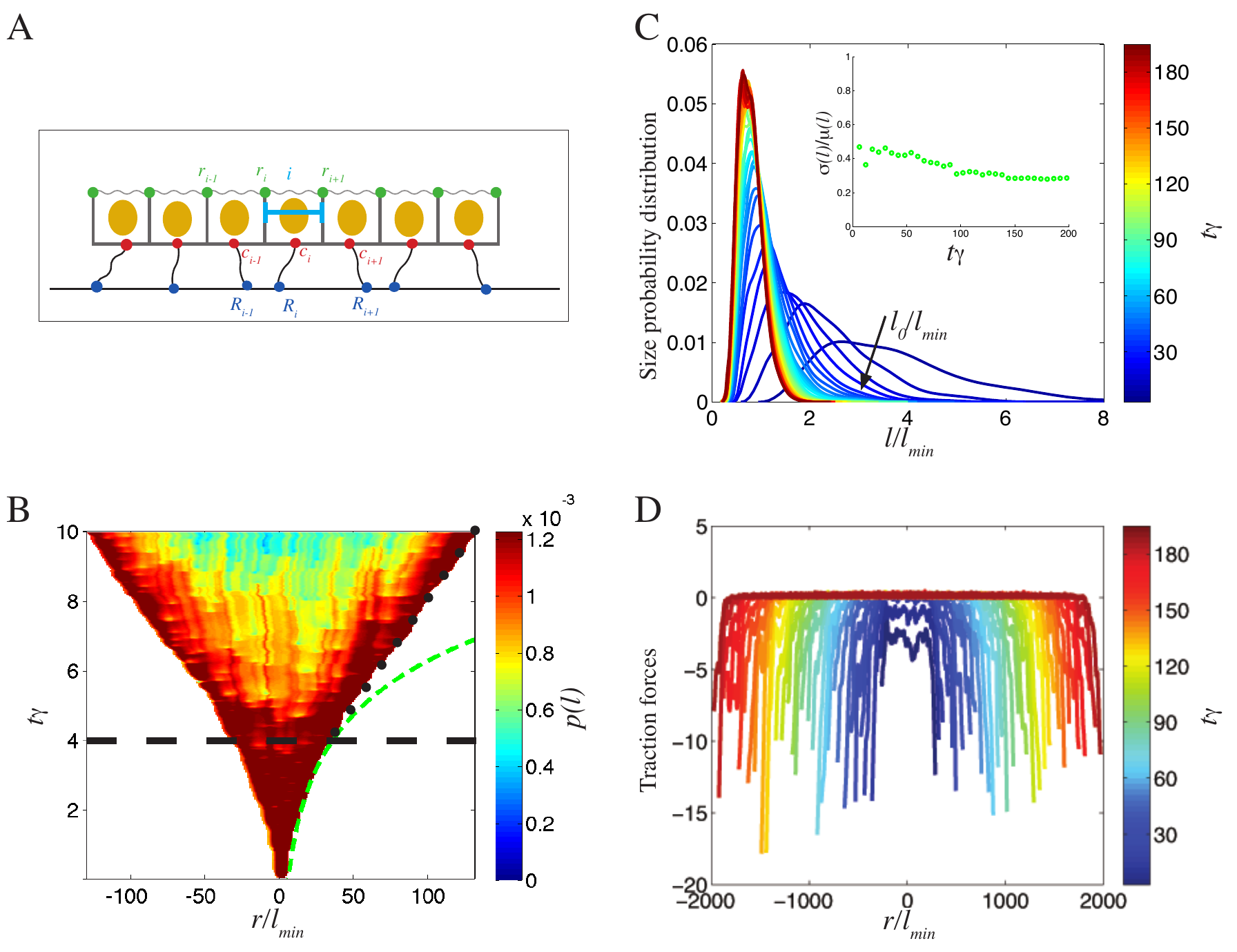}}
{\caption{{\bf Simulation results. }
{\bf A)} Sketch of the one-dimensional tissue growth model. Green springs represent cell elasticity, cell boundaries are marked in gray and cell attachments are represented as black tethers.
{\bf B)} Spatio-temporal profile of proliferation rate (indicated by the color) in the colony.  Initially, proliferation is uniform  and the colony size increases exponentially (dashed line) with time. At later times, proliferation in the bulk slows down and stops; in fixed size marginal zones rapid cell proliferation continues leading to a linear increase of the colony size (dotted line). 
{\bf C)} Cell size distribution as a function of time (coded by color).  The initial distribution around $l_0$ (set by the ratio of the rates of cell growth and proliferation) becomes broader and converges with time to a stationary distribution with mean below $l_{min}$. Inset shows the coefficient of variation.
{\bf D)} Traction force distribution throughout the colony at different times (coded by color). Note that small colonies are under tension. At later times only the margins of the colony are under tension, while  the center is stress free.
}}
\label{fig:model}
\end{figure*} 

\end{document}


\renewcommand{\thefigure}{SI.\arabic{figure}}
\renewcommand{\themovie}{SI.\arabic{movie}}
\renewcommand{\theequation}{SI.\arabic{equation}}

\maketitle


\section{Supplementary data}
\begin{figure}[h!]
\label{fig:si}
\centerline{\includegraphics[width=8cm]{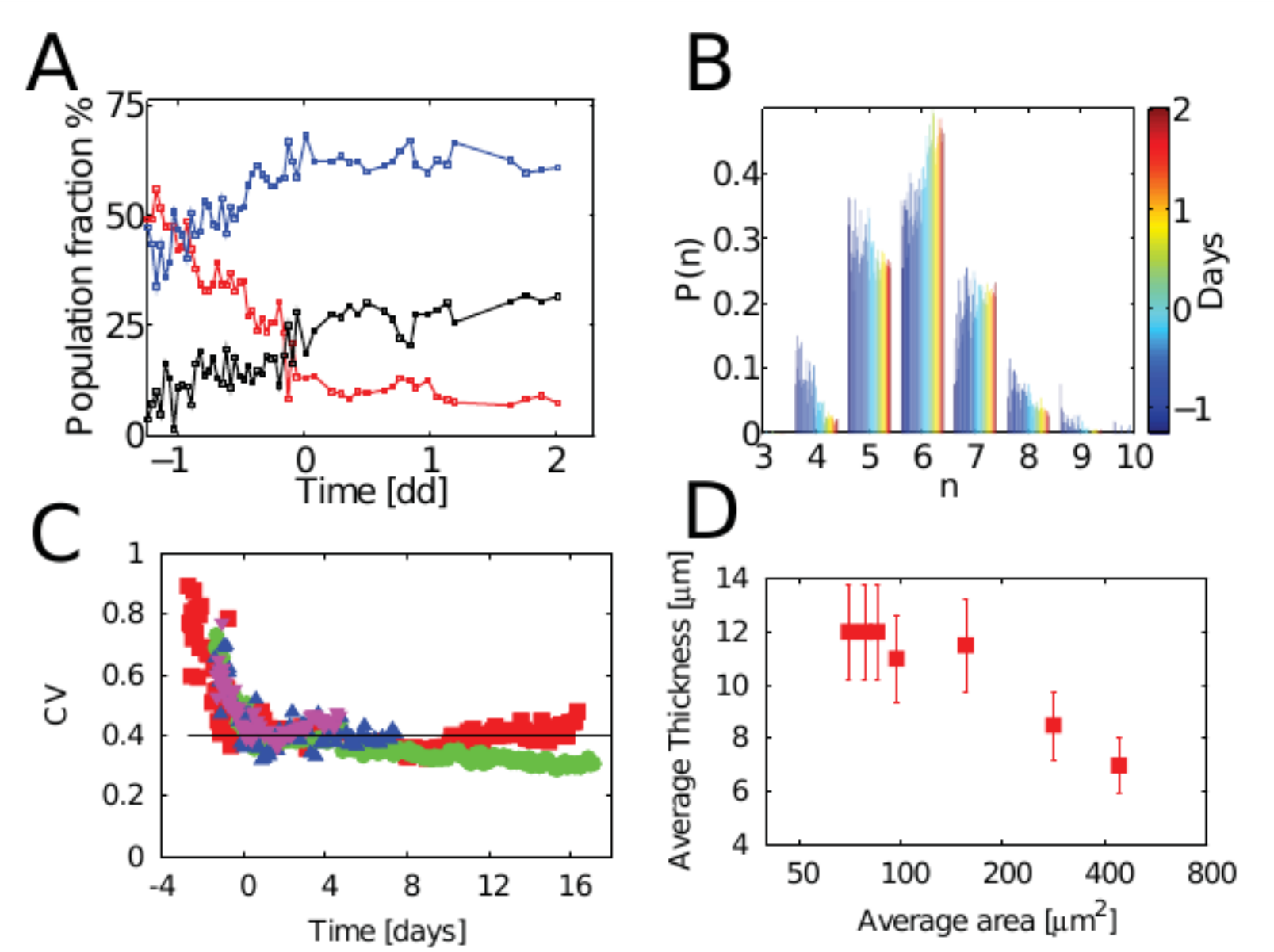}}
      \caption{{\bf A)} Aspect ratio of the Voronoi cells constructed for the segmented
        nuclei. Aspect ratio is defined as the ratio between the
        largest and the smallest eigenvalues of the inertia tensor
        defined by cell vertices. Regular polygons have aspect ratio
        between 1 and 1.5. The figure presents the fraction of the
        cell population with aspect ratio $ar< 1.5$ (black),
        $1.5<ar<3$ (blue) and $ar>3$ (red). As a function of
        time, the fraction of regular polygons increases and the fraction of
        deformed cells decreases dramatically. Time zero refers to the
        morphological transition.
        {\bf B)} Change in cellular the coordination number  across the morphological
        transition. Each histogram (coded by a color) represents the distribution
        of the number of cell vertices at a given time during the
        experiment. 
        The last distribution plotted is stationary and is analogous
        to the one found in other biological
        tissues~\cite{Lewis1928,Rivier1995,Gibson2006,Nagpal2008}. 
        {\bf C)} Cell size heterogeneity across the morphological
        transition. Different colors represent the coefficient of
        variation as a function of time for different experiments.
        The coefficient of variation was obtained by calculating the
        ratio of the difference of the quartiles 75 and 25 to the
        median: $(Q_{75}-Q_{25})/Q_{50}$. Data from different experiments were
        time aligned referring to the morphological transition.
        {\bf D)} Average cell thickness as a function of cell area. We observe that below 200$\mu$m, cell height does not
        change appreciably. Cell height is measured by finding the
        best focal plane at a given time and then tracking the
        position of the objective during the whole experiment.}
  \end{figure}

 \begin{movie}
    \caption{Dynamics of a growing epithelial colony. The movie is
      composed of 9x9 patches for a total of 3,02x4,05 mm. 6 cells were seeded and
      imaged for 10 days. Media were replaced daily. Scale bar: 1mm.}
  \end{movie}
  \begin{movie}
    \caption{Dynamics of epithelial tissue in the bulk. Cells were seeded at
      uniform density and imaged continuosly for a week. Media were replaced
      daily. Confluency is reached at $t=0.6$ days. By day 2 cell
      motion has completely disappeared, and only small scale
      vibrations can be observed. Scale bar: 100$\mu$m.}
  \end{movie}
\clearpage
\newpage
\section{Langevin model for tissue motility.}
To illustrate our interpretation of the observed interplay between cellular growth, motility and colony expansion, we formulate and analyze a simple growth model for adherent epithelial tissues. We choose as a point of departure a one dimensional version of the "vertex model"  as shown in Fig. 5A. Physical position of cell $i$ is specified by its vertices $r_i$ and $r_{i+1}$ which it shares with its neighbors. In addition to the interaction between neighboring cells we also include adhesion to the substrate. The attachment of cells to the extracellular matrix is mediated by many focal adhesions on the basal membrane of the cell. In our model we represent the cumulated effect of the focal adhesions by a single attachment point $R_i$ for each cell. We assume the mechanical properties of the tissue to be elastic on short time scales so that for a given set of intrinsic cell lengths and attachment points, the vertices of the cells are determined by minimizing the energy 
\begin{equation}\label{eq:vertx}
H(r_1, ..., r_{N+1}) = k \sum_{i=1}^{N} (r_{i+1}-r_i-L_i(t))^2 +  \kappa \sum_{i=1}^{N}[R_i -(r_{i+1}+r_i)/2]^2,
\end{equation}
where $L_i(t)$  is the intrinsic preferred length of the cell. The first term describes the mechanical interaction between cells, the second term accounts for the attachment of the cells to the substrate and the units are chosen so as to make attachment stiffness unity. In principle, depending on the mechanical stress $s_i \propto k (r_{i+1}-r_i-L_i(t))$  a cell can adapt its intrinsic length scale $L_i$ resulting in an effective plasticity of the tissue that would relax the stress at long times. Here, for simplicity, we will only include the effect of cell growth, which also manifests itself as a change - an increase - of $L_i$. We assume that $L_i$ resists compression $s_i<0$ and grows with rate $\alpha$ for $s_i>0$, e.g. $\frac{d}{dt} L_i=\alpha \theta(s_i)$, where $\theta$ is the step function. This implements the assumption that cell grow in size only when under tension, with the consequence in the absence of tension, cell divisions will reduce intrinsic cell size $L_i$.

The dynamics of the attachment points is driven by relaxation of elastic stress and by random forces generating cell motility
\begin{equation}\label{eq:langevin}
\sigma {d \over dt} R_i = - {\partial H \over \partial R_i } + \eta_i (t),
\end{equation}
where $\eta_i$ - a Langevin-type random force representing motility - is a Gaussian, white random function of time defined by its second moment $ < \eta_i (t) \eta_j (0) > = \Gamma \delta (t) \delta_{ij}$. The relevant time scale for the attachment dynamics is given by  $\sigma$ - which acts like friction. While we assume that the random force representing cell motility has a zero average in the bulk of the tissue, to represent the observed directed (outward) crawling of cells on the boundary, we allow $<\eta_N>=- \ <\eta_1> = \sigma v_{max}$, where $v_{max}$ sets the maximal crawling velocity of boundary cells. 

In addition to the continuous changes in cell sizes and attachments, cells may divide. Motivated by our experimental quantification of the cell area growth curve (Fig.~4) we make the average rate for cell division explicitly dependent on cell size $l_i=r_{i+1}-r_i$ by using $p(l_i)={\rm max}(\gamma(l_i-l_{\rm min}),0)$ in our simulations,   which is also consistent with our finding of a growth size check point. The division process replaces the dividing cell by two equivalent daughter cells with attachment points set to the middle of each cell and the sum of internal length matches the internal length of the mother cell.

We have simulated the dynamics of this model (with parameters $k=1, \kappa=.5, \alpha=.02 , \sigma=1, \Gamma=1,\gamma=.2 $) using custom-written Matlab programs implementing matrix inversion for the dynamics of the vertices and the Runge-Kutta method for the dynamics of the attachment-points. Programs are available upon request.

\section{ Additional information on the methods}
{\bf Cell culture}
Cells were seeded at uniform density (around 600 cells/mm$^2$) on a fibronectin (Sigma-Aldrich, F1141-2MG) coated PDMS membrane (McMaster-Carr, 87315K62) 
and imaged in phenol red free IMEM (Cellgro, 10-26-CV) supplemented with Penicillin-Streptomycin and 5\% FBS. The media was replaced daily and the culture conditions were kept at 37$^\circ$C and 5\% CO$_2$ by means of a custom made microscope 
stage enclosure.
Single colony experiments were performed by 
seeding cells at a density of about 1 cell/cm$^2$ 
in a glass bottom petri dish. Images of the colony spanning 9x9 contiguous 
fields of view 
were captured and stitched together, making it possible to 
follow single cell 
motion as well as tissue wide dynamics with good resolution. 

{\bf Image processing.}
Quantitative data on cell area was obtained 
with the help of contrast enhancement by Gabor-filtering~\cite{Lee1990,Jain1991}
and successive removal of low contrast regions. Frames taken 10 minutes apart were 
compared to remove poorly segmented cells. Cell size, shape, topology 
and structure functions were obtained from the segmented images.
Mean displacement measurements were 
made by cell tracking 
using a PIV-type analysis~\cite{Raffel1998}. 
The position of nuclei 
in a segmented image was propagated
in time by 
correlation analysis. 
The 
trajectory thus obtained was then corrected for 
stage drift 
and rms velocities were calculated as $(<v^2> - <v>^2)^{1/2}$. To extract the correlation length, 
the two point correlation function $C_v(R)$ for $\vec v-<\vec v>$ was computed and 
the length for which $C_v(R)=0.3$ was used.

{\bf Rate analysis}
To calculate the division rates shown in fig.~4E, 
eq.~1 was discretized; an over-determined linear system was constructed 
based on measurements of the size distribution at different times, and was subsequently pseudo-inverted~\cite{Ben-Israel2003}. The total number of cells per frame was obtained from the average area.